\documentclass[a4paper,11pt]{article}
\usepackage{pos}
\usepackage{multicol}
\title{Elastic $\pi-N$ scattering in the $I=3/2$ channel}

\author[a,b]{Constantia Alexandrou}
\author[a,b]{Kyriakos Hadjiannakou}
\author[b]{Giannis Koutsou}
\author[c]{Srijit Paul}
\author*[b]{Ferenc Pittler}
\author[d]{Marcus Petschlies}
\author[a,d,e]{Antonino Todaro}

\affiliation[a]{Department of Physics, University of Cyprus, 20537 Nicosia, Cyprus}
\affiliation[b]{Computation-based Science and Technology Research Center, The Cyprus Institute, Cyprus}
\affiliation[c]{Institut f¨ur Kernphysik, Johannes Gutenberg-Universit¨at Mainz, 55099 Mainz, Germany}
\affiliation[d]{HISKP (Theory), Rheinische Friedrich-Wilhelms-Universit\"at Bonn, Germany}
\affiliation[e]{Dipartimento di Fisica and INFN, Universit\'a di Roma ``Tor Vergata'', I-00133 Rome, Italy}
\affiliation[f]{Institut f\"ur Physik, Humboldt-Universit\"at zu Berlin, Newtonstrasse 15, 12489 Berlin, Germany}

\emailAdd{f.pittler@cyi.ac.cy}

\abstract{ We present our study of $\pi-N$ scattering in the iso-spin $I=3/2$ channel 
for the first time at the physical point. The calculation is performed using $N_f=2+1+1$ flavors of 
twisted mass fermions with clover improvement at physical pion mass. We compute energy levels for the
rest frame and moving frames up to a total momentum of $|\vec{P}|=\sqrt{3} \,\frac{2\pi}{L}$,
and for all the relevant ireducible representations of the lattice symmetry groups. We perform a 
phase-shift analysis including $s\,(\ell=0)$ and $p\,(\ell=1)$ wave phase shifts assuming a Breit-Wigner 
form and determine the parameters of the $\Delta$ resonance.
}

\FullConference{%
 The 38th International Symposium on Lattice Field Theory, LATTICE2021
  26th-30th July, 2021
  Zoom/Gather@Massachusetts Institute of Technology
}

\begin{document}
\maketitle

\section{Introduction} 


The study of resonances with lattice methods in Euclidean field theory has matured from the initial proposal by L\"uscher
\cite{Luscher:1990ux} to deploy the finite volume as a probe for scattering properties of hadrons.
A prime example that speaks to these current capabilities are the numerous studies of meson-meson scattering,
and in particular of the $\rho(770)$ meson. Determinations of $\rho$ resonance parameters have been carried 
out by a number of lattice QCD collaborations. More recently, lattice calculations of the $\rho$ have been advanced even to the
point of physical pion mass ~\cite{Fischer:2020yvw,ExtendedTwistedMass:2019omo,Akahoshi:2021sxc,Guo:2016zos,Andersen:2018mau}.


With meson-baryon scattering the situation is less advanced by comparison. 
A natural step is the study of the simplest resonance in pion-nucleon ($\pi-N$) scattering, namely the well-known $\Delta(1232)$. 
Moreover, with presently available gauge configurations we can approach elastic $\pi-N$ scattering at physical pion mass,
with realistic comparison to experimental results.
This is the topic of our contribution. The $\Delta$ is the lightest baryon spin-3/2 resonance. It is well isolated from other resonances
and decays almost exclusively to $\pi-N$. The decay $\Delta\rightarrow N\gamma$ corresponds to less than 1\% of the
total decay width. Besides the theoretical interest, the presence of $\Delta$  in compact neutron stars makes
it important for phenomenology as well. The interaction between $\Delta$ and nucleon could generate thermodynamic
instabilities in compact neutron stars \cite{Raduta:2021xiz}.

\begin{figure}[htpb]
  \centering
  \includegraphics[width=0.45\textwidth]{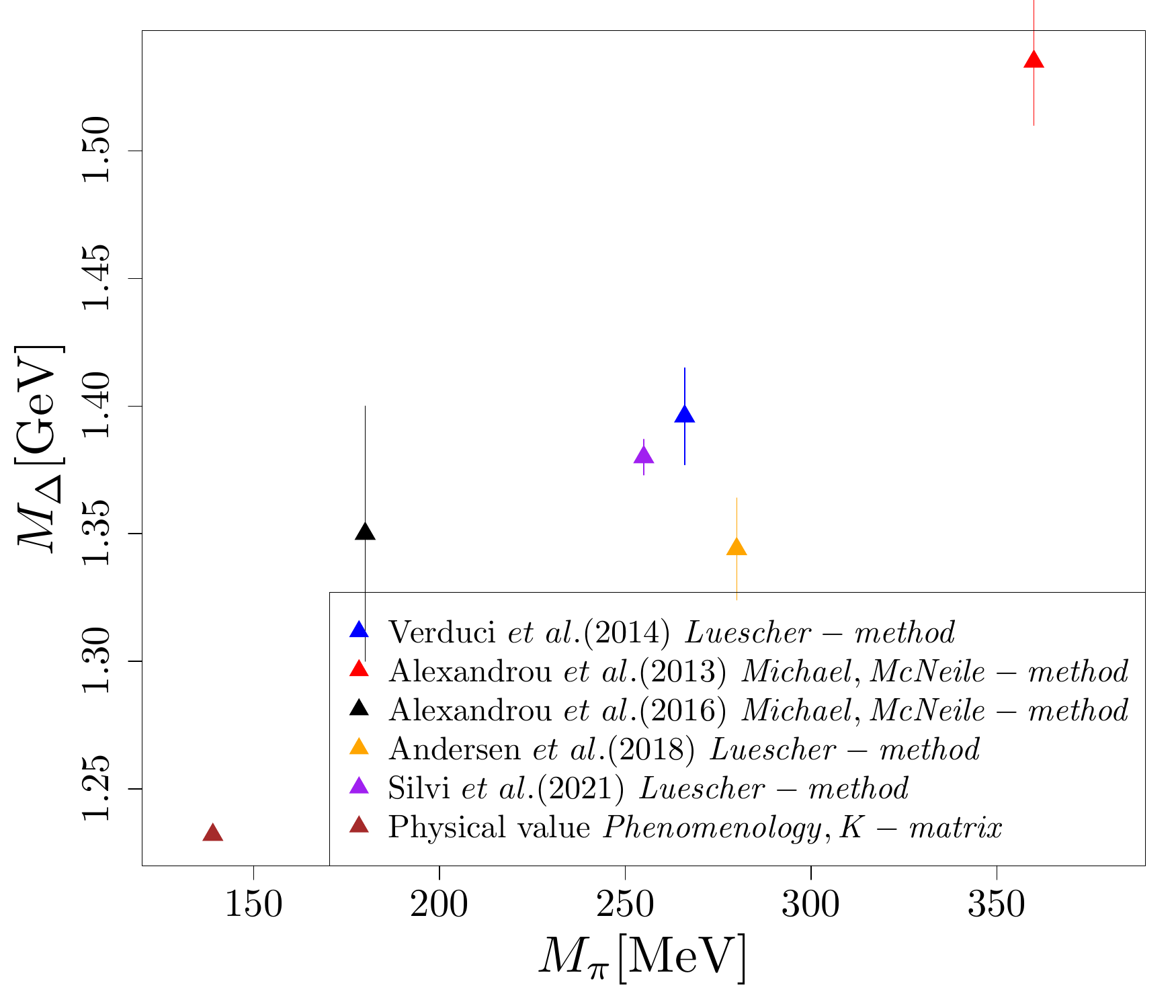}
  \includegraphics[width=0.45\textwidth]{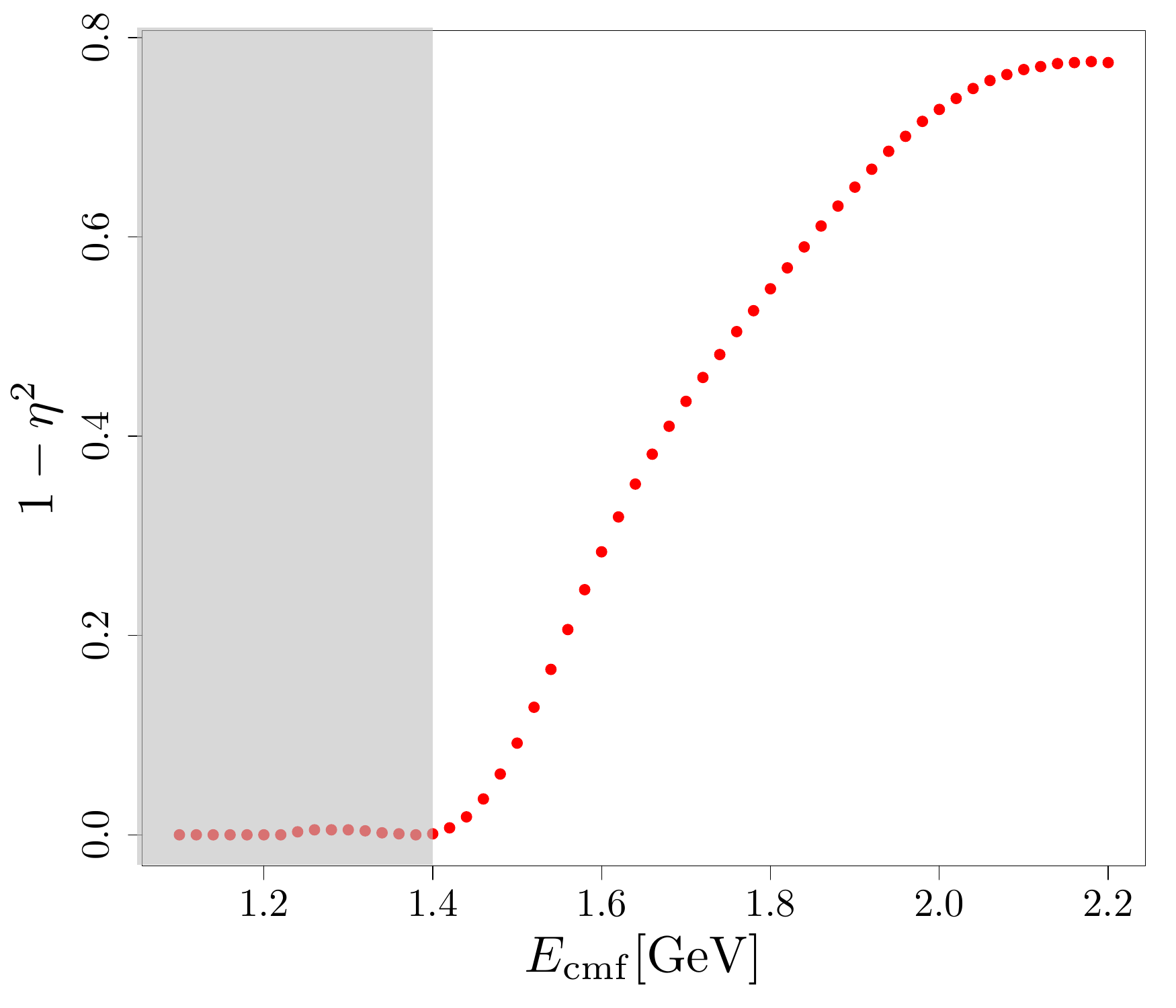}
  \caption{Left: Previous determinations of the $\Delta$-resonance mass as function of the pion mass. In italics we give the method which is used 
    in the computation.
    Right: The inelasticity of the scattering in the $\Delta$ channel, shaded gray area shows the energy range considered in this work. Data is taken from \cite{datacenter}.}
  \label{fig:history}
\end{figure}

With the left-hand plot in Fig.~\ref{fig:history} we briefly comment on already available lattice studies of the $\Delta$ and its coupling to the $\pi-N$ state.
Two different type of methods were previously used to study $\Delta$ resonance parameters:
the L\"uscher approach mentioned above and denoted as \textit{Luescher-method} in the legend, 
and the transfer matrix  approach introduced in Ref.~\cite{McNeile:2002az} denoted as \textit{Michael,McNeile-method}.
However, the latter method assumes a quasi-stable $\Delta$ state perturbatively close in mass to $\pi-N$ threshold, to extract
the $\Delta - N\pi$ transition matrix element and the coupling. 
At this early stage the displayed individual studies given by the pion mass, at which they were carried out, differ completely by lattice action, 
type and size of discretization artifacts and finite-volume effects.
We report on our L\"uscher-method based $\pi-N$ elastic scattering analysis in the $\Delta$ channel,
for the first time directly at physical pion mass and with strange and charm quark close to their physical values.
This setup provides us with the well-known numeric difficulty of fast exponential deterioration of the signal-to-noise ratio in 
(meson-)baryon correlators and given the low pion mass only a small energy range of the $\pi-N$ system 
for elastic scattering $E_{2,\mathrm{thr}} = m_N + m_{\pi} \le \sqrt{s} \le m_N + 2 m_{\pi} = E_{3,\mathrm{thr}}$, between 2- and 3-particle threshold,
where the 2-particle quantization condition is rigorously applicable. However, we take the SAID inelasticity data \cite{datacenter} 
shown in the right-hand plot of Fig.~\ref{fig:history} as indication,
that $p$-wave $\pi-N$ scattering can be treated as elastic up to $\sqrt{s} = 1.4~\mathrm{GeV}$.


\section{Simulation details}
In our simulation setup we use $N_f=2+1+1$ flavors of twisted-clover fermions with physical quark masses from the Extended Twisted Mass Collaboration (ETMC).
The important ensemble properties regarding this work are summarized in Table~\ref{tab:ensemble_params}. 
More detailed description is given in Ref.~\cite{Alexandrou:2018egz}. 
\begin{table}
\begin{center}
\begin{tabular}[htb]{c|cccc|cc}
ensemble &  $M_{\pi} / \mathrm{MeV}$  & $M_N / \mathrm{MeV}$  & $M_\pi L$ & $a / \mathrm{fm}$ & $N_{\mathrm{conf}}$ & $N_{\mathrm{src}}$ \\ 
\hline
cB211.072.64 & $139.43(9)$ & $944(10)$  & $3.622(3)$ & $0.0801(2)$ & 388 & 64 \\ 
\hline
\end{tabular}
\caption{\label{tab:ensemble_params} Parameters of the ensemble used in this work; further details are given in Ref.~\cite{Alexandrou:2018egz}.
The right-most columns give the number of gauge configurations employed $N_{\mathrm{conf}}$ and the number of point sources
used per configuration $N_{\mathrm{src}}$.
}
\end{center}
\end{table}

As an input for the L\"uscher analysis we determine the spectrum of the $\pi-N$ system
in the $I=3/2$ channel from correlation functions of single- ($\Delta$-like ) and 2-hadron ($\pi N$ ) 
interpolators
\begin{align*}
  C_{XY}(\vec{p},t) &= \langle O^{\vec{p}}_{X}(t)\,\,{\bar{O}}^{\vec{p}}_{Y}(0) \rangle \,,\quad 
X,\,Y \, \in \, \lbrace \Delta, \pi N\rbrace \,,
\end{align*}
with total momentum (and individual particle momenta)
up to $|\vec{p}|=\sqrt{3}\,\frac{2\pi}{L}$.
%
%
We use the standard interpolating operators for the $\Delta$ at maximal $I_3 = +3/2$, the $\pi^+$ and proton $N$ given by:
\begin{equation}
  O_{\Delta, \mu}=\varepsilon_{abc} \, \left(u^a \, C\gamma_\mu \,  u^b\right) \, u^c\,,\quad
O_\pi = {\bar d}\, \gamma_5 \, u \,, \quad
O_N=\varepsilon_{abc} \, \left(u^{at} \, C\gamma_5 \, d^b\right) \, u^c
\end{equation}
For the $\pi-N$ 2-hadron interpolator we use the product of the single hadron interpolators $\pi$ and $N$.
Here we only briefly review our method for calculating two hadron two-point functions.
For further details we refer to Ref.~\cite{Silvi:2021uya}. Describing the propagation from source to sink we use point source propagators and in the case of the two-hadrons 
we use the sequential source technique. The sink to sink propagation is replaced in the $\pi N$ to $\pi N$ correlation functions by  fully time diluted stochastic source/propagator pairs, along which we cut the diagrams into reusable ``factors''. For the latter we implemented GPU-kernels, which reduce 2- and 3-fold propagator products, including momentum projection. All propagators receive Gaussian smearing at source and sink, with $N_G = 140$ smearing steps
and weight $\alpha_G = 0.5$, with APE smeared gauge field in the Gaussian smearing kernel with $N_{APE} = 60 $, $\alpha_{\mathrm{APE}} = 0.5$.

In discretized and finite volume we project interpolating operators to the irreducible representations (irreps $\Lambda$ with rows $\mu$) 
of the lattice rotation
group $^2O_h$ in the rest frame of the $\pi-N$ system, and the irreps of its subgroups in moving frames with total momentum
$\vec{p}_{tot}$ (little groups $LG(\vec{p}_{tot})$ ).
The subduction coefficients $U$ for the group projection of correlators
\begin{equation}
C^{\Lambda,\beta,\mu,\vec{p}_{tot}}_{{\mathcal A}_{\mathrm{sink}},{\mathcal A}_{\mathrm{source}}}=
\sum_{{\mathcal B}_{\mathrm{sink}},{\mathcal B}_{\mathrm{source}}}U^{\star,\beta\mu o_{\mathrm{sink}}}_{{\mathcal B}_{\mathrm{sink}}}
U^{\beta\mu o_{\mathrm{source}}}_{{\mathcal B}_{\mathrm{source}}}C_{{\mathcal B}_{\mathrm{sink}}{\mathcal B}_{\mathrm{source}}}
\end{equation}
with compound indices ${\mathcal B}=\lbrace\vec{p}_{\pi},\vec{p}_{N},\alpha\rbrace$
for particle momenta and irrep occurrence,
are determined based on \cite{Gockeler:2012yj} with a Gram-Schmidt decomposition for multiple occurrences.
The sum at the source and sink is restricted to have a fixed momentum amplitude i.e. 
${\mathcal A}=\lbrace {\vert\vec{p}\vert}^2_{\mathrm{\pi}},{\vert\vec{p}\vert}^2_{\mathrm{N}},o\rbrace$. 
To obtain the first few energy levels
we solve the generalized eigenvalue problem (GEVP) for the correlation matrix ($C^{\Lambda,\beta=0,\mu,\vec{p}_{tot}}$) 
and fit the resulting eigenstates with a single exponential decay:
\begin{equation}
  C_{ij}^{\Lambda,\mu,\vec{p}_{\mathrm{tot}}}(t)\, u_j^n(t)=\lambda^n(t,t_0) \, C_{ij}^{\Lambda,\mu,\vec{p}_{\mathrm{tot}}}(t_0) \, u_j^n(t_0)
\end{equation}
\begin{equation}
\lambda^n(t,t_0)\propto e^{-E_n\left(t-t_0\right)}.
\end{equation}

We summarize in Table~\ref{tab:irrep_this_work} the momenta and irreps used in this work together with the dominant partial waves the given irreps contribute to, 
and the size of the correlation matrix used in the GEVP analysis.
\begin{table}
\caption{\label{tab:irrep_this_work} Momentum frames and irreducible representations used in this work. Second column ($\ell$) indicates the infinite
volume partial waves contributing to the irreducible representation. In the third column we show the size of the 2d matrix used in the
GEVP for the particular irrep.
}

\begin{center}
\setlength{\tabcolsep}{10pt}
\renewcommand{\arraystretch}{1.2}
\begin{tabular}{|llll|}
\hline
$\vec{p}_{\mathrm{tot}} / (2\pi/L)$ & $\Lambda$ & $\ell$ & $N_{dim}$  \\
\hline
$(0,0,0)$ & $\mathrm{G1u}$ & $s,\ldots$     & 8x8 \\
$(0,0,0)$ & $\mathrm{Hg}$   & $p,f,\ldots$   & 9x9 \\
$(0,0,1)$ & $\mathrm{G1}$   & $s,p,d\ldots$  & 24x24 \\
$(0,0,1)$ & $\mathrm{G2}$   & $p,d,\ldots$   & 18x18 \\
$(1,1,0)$ & $\mathrm{(2)G}$ & $s,p,d,\ldots$ & 30x30 \\
$(1,1,1)$ & $\mathrm{(3)G}$ & $s,p,d,\ldots$ & 16x16 \\
$(1,1,1)$ & $\mathrm{F1}$   & $p,d,\ldots$   & 6x6 \\
$(1,1,1)$ & $\mathrm{F2}$   & $p,d,\ldots$   & 6x6 \\
\hline
\end{tabular}
\end{center}
\end{table}

\section{Determining the interacting spectrum}

\begin{figure}[htpb]
  \centering
  \includegraphics[width=0.47\textwidth]{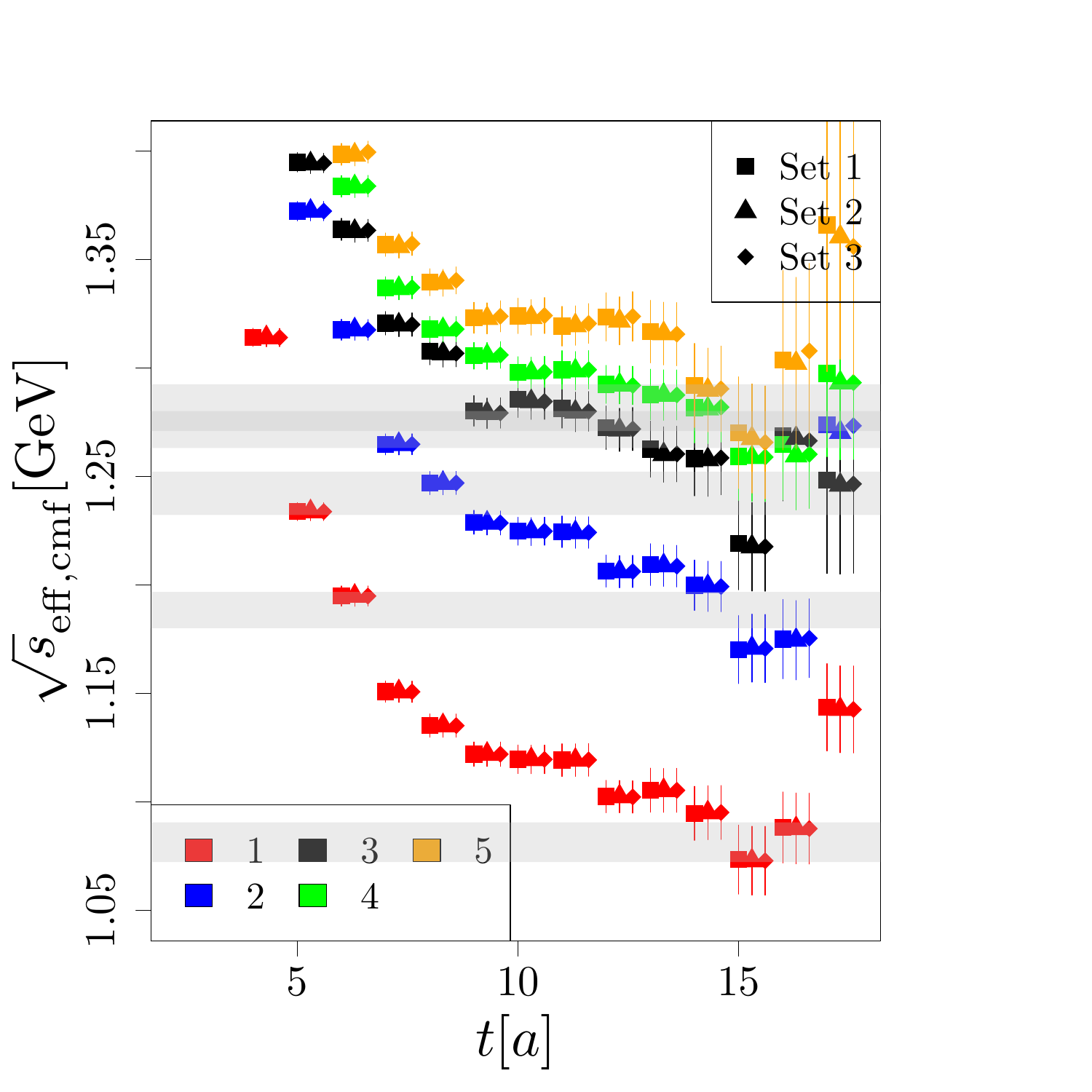}
  \includegraphics[width=0.47\textwidth]{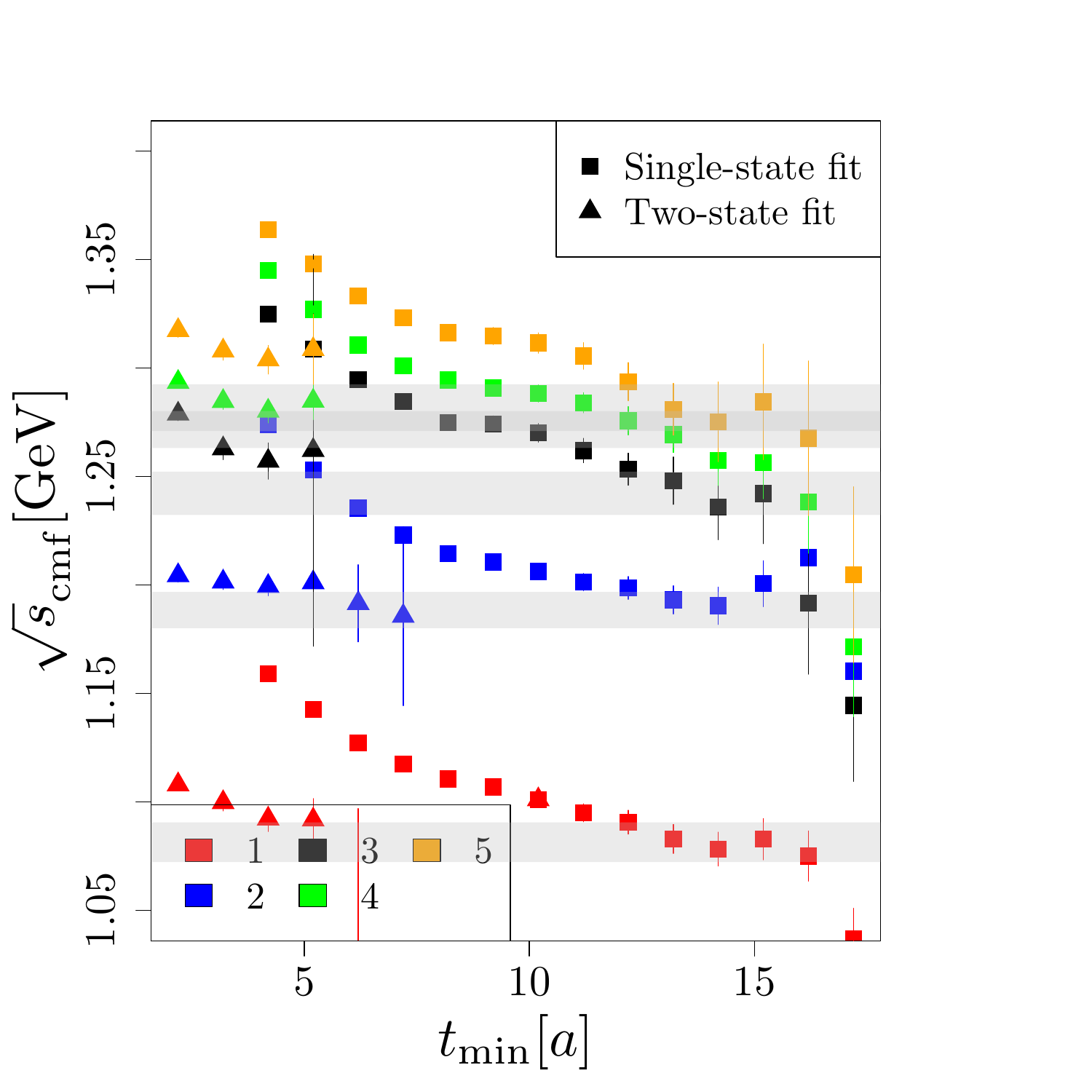}
  \caption{Example for GEVP results in irrep $G_1$ of the little group $2C_{4\mathrm{v}}$. Left: 
The dependence of the principal correlator effective mass on the basis for the GEVP.  $\mathrm{Set~1}$ 
consists of interpolators from all available momentum configurations and occurrences, $\mathrm{Set~1} =  \lbrace \Delta_{1-4};\pi_1N_0\vert_{1-2};
\pi_0N_1\vert_{1-2},\pi_2N_1\vert_{1-4};\pi_1N_2\vert_{1-4};\pi_3N_2\vert_{1-4};\pi_2N_3\vert_{1-4}\rbrace$, where
the subscript denotes particle momentum and occurrences $|\vec p|/(2\pi/L)\,|1-\cdots $.
In $\mathrm{Set~2}$ we take into account half of the occurrences from $\mathrm{Set~1}$ using the signal 
quality for selection criteria. In $\mathrm{Set~3}$ we have replaced the two occurrences from Set 2 from momentum combinations $\pi_2N_3$ and $\pi_3N_2$ with the other two in Set 1. 
Right: Stability of single and two states fits to the principal correlators as a function of $t_{\mathrm{min}}$.
Gray bands are representing the final fit results from the chosen fit range.}
\label{fig:gevp_details}
\end{figure}

The interaction in the spectrum shows up as a resulting shift of the energy levels with respect 
to the non-interacting ones. The energy shift is determined from a GEVP. The basis for the GEVP
were selected by requiring a stable behavior of the effective mass under removing/changing a few basis vectors. 
An example for the basis selection 
we show on the left plot of Fig.~\ref{fig:gevp_details}. We extract the energy levels using single
state fits to the principal correlators. We choose $t_\mathrm{min}$ by requiring agreement between 
single and two-states fits and stability by increasing or decreasing $t_{\mathrm{min}}$ by one in 
lattice units. An example for the stability we show on the right panel of Fig.~\ref{fig:gevp_details}.
the results from single and two-state fits.

\begin{figure}[]
\begin{center}
 \includegraphics[scale=0.55]{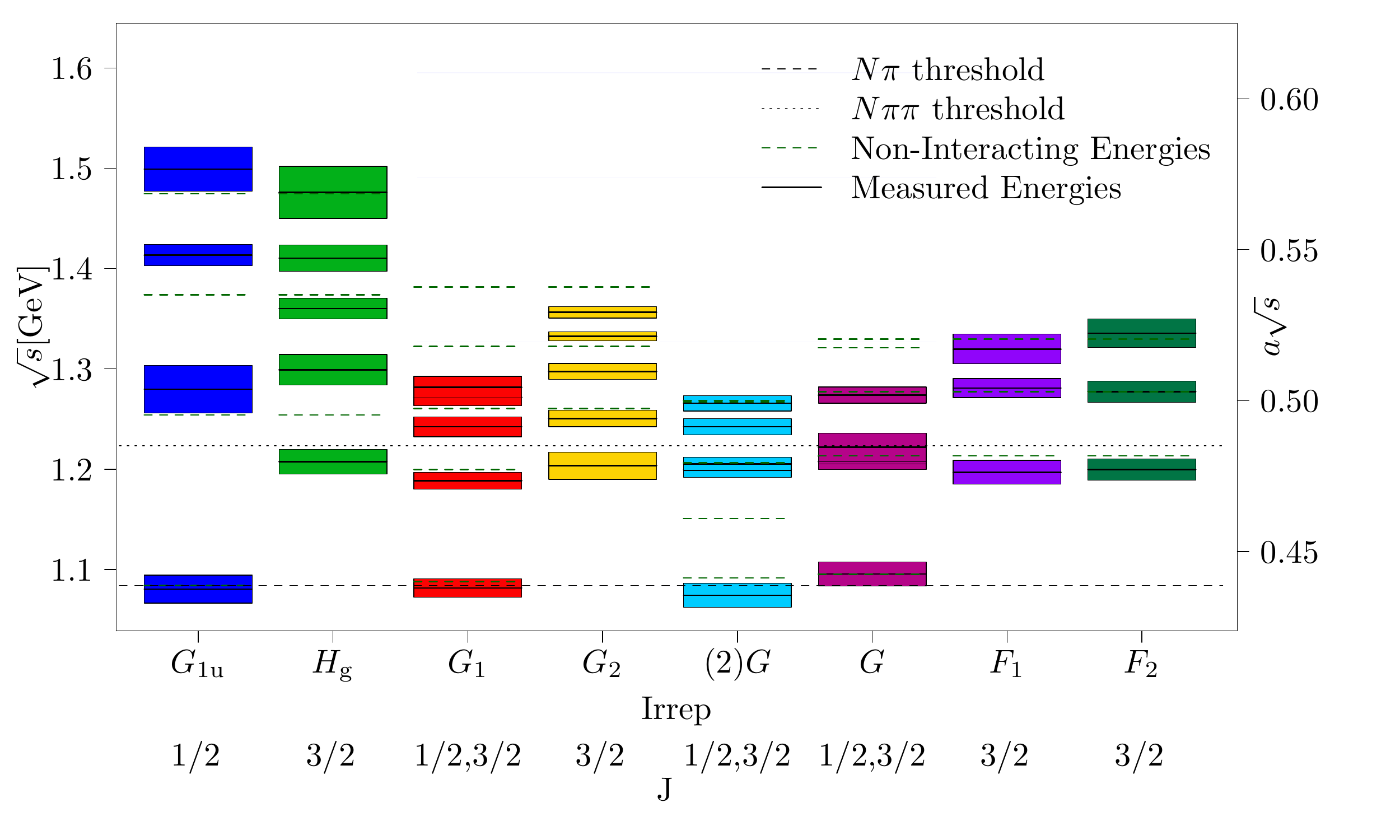}
\end{center}
 \caption{\label{fig_sprectrum_summary} The invariant mass $\sqrt{s}$ for the levels of the $\pi-N$ system in the rest and moving frames up to 3 lattice units of momentum and for all irreducible representations of the lattice rotation group. The $N\pi$ and $N\pi\pi$ thresholds are indicated by dashed and dotted horizontal lines respectively. The non-interacting energies for each irrep and frame is indicated by bold dashed horizontal segment.}
\end{figure}

Our results for the interacting spectrum are shown in Fig.~\ref{fig_sprectrum_summary}. As an input for the L\"uscher-analysis we 
restrict the energy window to be below 1.3 GeV, in order to keep contamination from inelastic scattering small.

\section{Phase shift analysis}

The connection between the finite volume two-particle interacting spectrum and the infinite-volume resonance parameters is 
encoded in the L\"uscher quantization conditions (LQC-s). Mathematically, they are determinant equations given by
\begin{equation}
\label{eq:lqc}
\mathrm{det}\left(M^{\Lambda}_{J\ell n,J^{\prime}\ell^{\prime}n^{\prime}}(s)-\delta_{JJ^{\prime}}\,
\delta_{\ell\ell^{\prime}}\,
\delta_{nn^{\prime}}\,\,
\mathrm{cot}\delta_{J\ell}(s)\right)=0,
\end{equation}
where $M^\Lambda$ is the finite volume L\"uscher function\cite{Gockeler:2012yj}, $\delta_{J\ell}$  is the infinite volume scattering phase shift for
total angular momentum $J$ and orbital angular momentum $\ell$, and $n$ is the occurrence of the irreducible representation $\Lambda$. 
We parameterize the energy dependence of  the infinite volume phase shift. In this work, we use the $s$ and $p-$wave phase shift parameterized by a 
constant scattering length and a Breit-Wigner type of resonance, respectively.
\begin{align}
\mathrm{cot} \, \delta_{\frac{1}{2}0}(s) &= a_0\, q_{\mathrm{cmf}}(s)\,, 
\quad  q_{\mathrm{cmf}}^2(s) = \frac{\left(s-M_N^2-M_\pi^2\right)^2-4M_N^2M_\pi^2}{4s} 
\label{eq:delta10}\\
\mathrm{tan}\, \delta_{\frac{3}{2}1}(s) &= \frac{\sqrt{s} \, \Gamma(\Gamma_R,M_R,s)}{M_R^2-s}\,,
\quad \Gamma(\Gamma_R,M_R,s)=\Gamma_R\left(\frac{q_{\mathrm{cmf}}(s)}{q_{\mathrm{cmf}}(M_R^2)}\right)^3\frac{M_R^2}{s}
\label{eq:delta32}
\end{align}
\begin{figure}
\begin{tabular}{cc}
\includegraphics[scale=0.35]{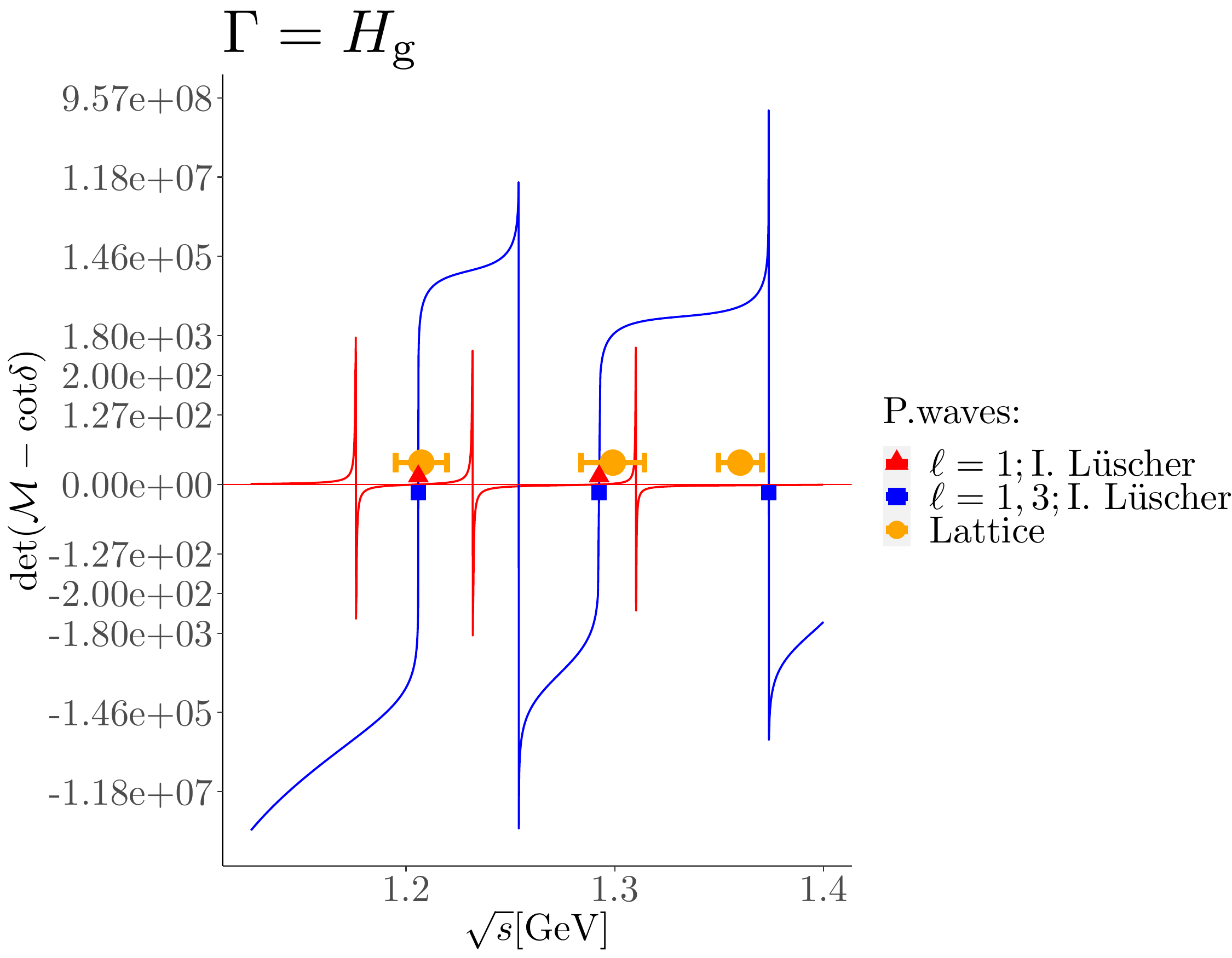}
\includegraphics[scale=0.35]{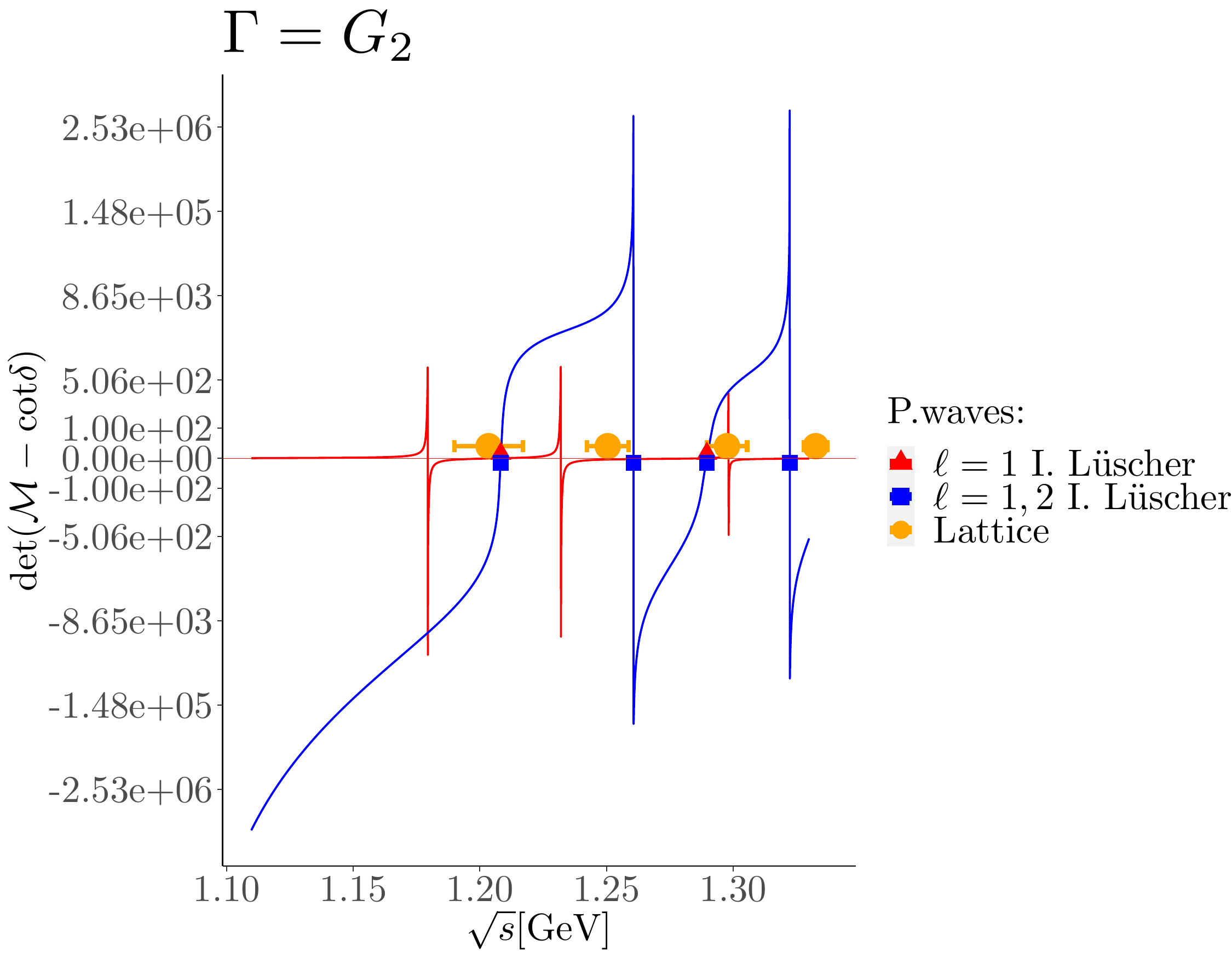}
\end{tabular}
\caption{\label{fig:QC_higher} LQC for irrep Hg (left) and G2 (right) using partial waves $\ell=1,2,3$. LQC with $\ell=1$ ($\ell\ge 1$) shown by continuous red (blue) curve and corresponding roots with triangle (rectangle) symbols. Lattice results are shown with big yellow circle with error bar. Red triangles and blue circles are obtained using the inverse L\"uscher method as described the text in detail.}
\end{figure}
\begin{figure}[]
\begin{center}
 \includegraphics[scale=0.4]{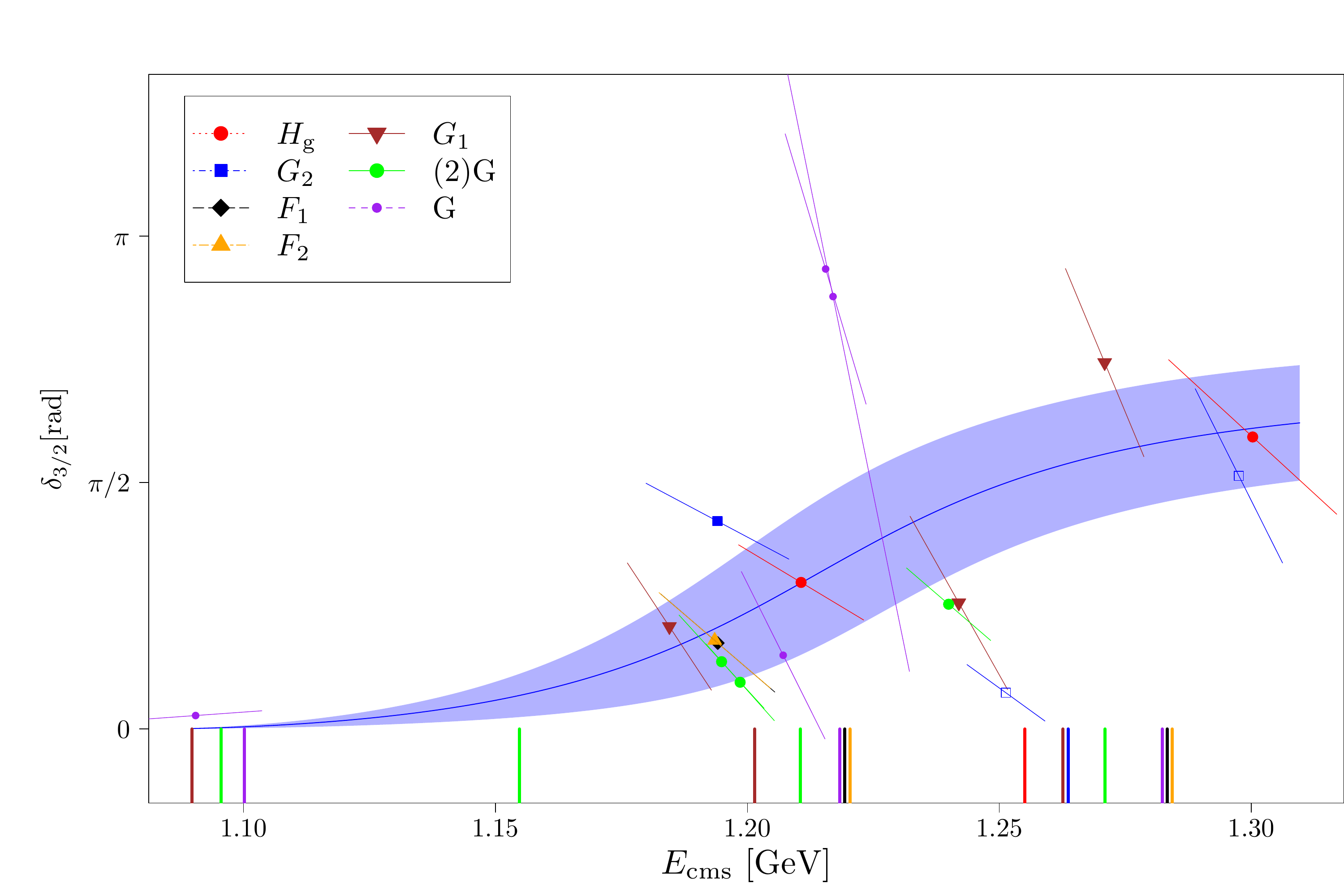}
\end{center}
 \caption{\label{fig:final}$I=3/2$ phase shift of the  $\Delta$ resonance. Errors are determined using jackknife resampling. At the bottom of the plot with thin vertical lines we indicate the non-interacting energies.}
\end{figure}
To obtain the L\"uscher prediction of the finite volume spectrum for a given resonance mass ($M_R$) and 
resonance width ($\Gamma_R$) we solve the LQC eq.~(\ref{eq:lqc}) numerically.
In the angular-momentum we truncate the LQC by $\ell=1$, and consider only the two partial waves in eq. (\ref{eq:delta10}) and (\ref{eq:delta32}).
To estimate the effect of the truncation in 
$\ell$
we determine the roots of LQC using physical parameters for $s$ and $p-$wave and including $\ell=3$ ($\ell=2$) in the center-off-mass 
frame (moving frames). We use $\mathrm{tan}\left(\delta_{J\ell}\right)=(a_{\ell}q_{\mathrm{cmf}})^{2\ell+1}$ to parameterize
higher partial waves with $a_\ell=-0.001\frac{1}{\mathrm{MeV^{2\ell+1}}}$. We show the comparison of this test with our numerically determined spectrum on 
Fig.~\ref{fig:QC_higher}. These results show that the third energy level in the Hg and already the second energy level in the G2 irrep
cannot be explained solely using the dominant $p$-wave approximation and thus are not expected to play a role in the description of the resonance. For this reason we omit them from the final analysis. The parameters of the $\Delta$ resonance are determined through a non-linear fit of the energy levels to the finite volume spectrum obtained by the LQC-s. For the parameters of the resonance we obtain $M_R=1255(25)~\mathrm{MeV},\Gamma_R=140(120)~\mathrm{MeV}$ and for the $s$-wave scattering length $a_0=-0.0016(6)\mathrm{MeV}^{-1}$ from the fit with an overall $\chi^2/\mathrm{dof.}=0.88$. 
We show the fitted phase-shift curve together with the computed phase-shifts for the energy levels in Fig.~\ref{fig:final}. The uncertainty band on the phase-shift curve is determined using jackknife resampling.




\section{Conclusions}
Our presented analysis is the first numerical resonance calculation at the physical 
point in the meson-baryon sector. We determined the parameters of the Delta resonance 
using L\"uscher's method and find
\begin{align}
  M_R &= 1255\,(25)~\mathrm{MeV} \,, \quad \Gamma_R = 140\,(120)~\mathrm{MeV} \qquad (m_{\pi} = 139.43(9) \,\mathrm{MeV})
  \label{eq:BW-results}
\end{align}
compared to the experimentally determined values of $M^{\mathrm{exp}}_R=1232$~MeV and $\Gamma^{\mathrm{exp}}_R=120$~MeV. 
Our analysis is barely sensitive to the width of the resonance.
Our future plans include the determination of the scattering length and investigation of the $I=1/2$
channel as well, in order to determine the $\sigma$-term.

\section{Acknowledgment}
\small
We would like to thank all members of ETMC for their constant and pleasant collaboration.
The project is supported by PRACE under project ``The $N\pi$ system using twisted mass fermions at the 
physical point'' (pr79), the measurements are done on the Piz-Daint cluster at CSCS.
KH is supported by the Cyprus Research Promotion foundation under contract number POST-DOC/0718/0100 and EuroCC project 
funded by the Deputy Ministry of Research, Innovation and Digital Policy and Cyprus The Research and Innovation Foundation and
the European High-Performance Computing Joint Undertaking (JU) under grant agreement No 951732. The JU receives
support from the European Union’s Horizon 2020 research and innovation program. FP acknowledges financial support from
the Cyprus Research and Innovation Foundation under project ''NextQCD``,
contract no. EXCELLENCE/0918/0129. 
MP gratefully acknowledges support by the Sino-German collaborative research center CRC-110.
The open source software packages
QUDA~\cite{Clark:2009wm,Babich:2011np,Clark:2016rdz}, 
R~\cite{R:2019,hadron:2020} have
been used.

\end{document}